\documentclass[
aps,%
12pt,%
final,%
notitlepage,%
oneside,%
onecolumn,%
nobibnotes,%
nofootinbib,%
superscriptaddress,%
noshowpacs,%
centertags]%
{revtex4}

\usepackage{cmap}
\usepackage{epstopdf}
\usepackage{graphicx}
\usepackage{longtable}

\begin{document}

\title{Infrared Morphology of Regions of Ionized Hydrogen} 
\author{A. P. Topchieva}
\email{stasyat@inasan.ru}
\affiliation{Institute of Astronomy, Russian Academy of Sciences, Moscow, 119017 Russia}

\author{D. S. Wiebe}
\email{dwiebe@inasan.ru}
\affiliation{Institute of Astronomy, Russian Academy of Sciences, Moscow, 119017 Russia}

\author{M. S. Kirsanova}
\email{kirsanova@inasan.ru}
\affiliation{Institute of Astronomy, Russian Academy of Sciences, Moscow, 119017 Russia} 
\affiliation{Institute of Astronomy, Russian Academy of Sciences, Moscow, 119017 Russia Barkhatova Kourovka Astronomical Observatory, Yeltsin Ural Federal University,Yekaterinburg, 620000 Russia}

\author{V. V. Krushinskii}
\email{vadim.krushinsky@urfu.ru}
\affiliation{Institute of Astronomy, Russian Academy of Sciences, Moscow, 119017 Russia Barkhatova Kourovka Astronomical Observatory, Yeltsin Ural Federal University,Yekaterinburg, 620000 Russia} 

\begin{abstract}
A search for infrared ring nebulae associated with regions of ionized hydrogen has been carried out. The New GPS Very Large Array survey at 20 cm forms the basis of the search, together with observations obtained with the Spitzer Space Telescope at 8 and 24 $\mu$m and the Herschel Space Telescope at 70 $\mu$m. Objects having ring-like morphologies at 8 $\mu$m and displaying extended emission at 20 cm were selected visually. Emission at 24 $\mu$m having the form of an inner ring or central peak is also observed in the selected objects. A catalog of 99 ring nebulae whose shapes at 8 and 70 $\mu$m are well approximated by ellipses has been compiled. The catalog contains 32 objects whose shapes are close to circular (eccentricities of the fitted ellipses at 8 $\mu$m no greater than 0.6, angular radius exceeding 20). These objects are promising for comparisons with the results of one-dimensional hydrodynamical simulations of expanding regions of ionized hydrogen.

\end{abstract}

\maketitle

\section{INTRODUCTION}\label{sec:intro}

Studies of the morphology of regions of infrared (IR) emission in the Galaxy have shown that this emission fairly often forms ring or arc structures, suggestive of the impact of hot stars on the interstellar medium \cite{1988ApJ...329L..93V}. The widespread presence of such structures became clear thanks to observations made with the Spitzer Space telescope ~\cite{2006ApJ...649..759C,2007ApJ...670..428C}. The catalog of such objects compiled using data from the Spitzer and WISE IR telescopes currently contains more than 5000 objects \cite{2012MNRAS.424.2442S}. These objects are sometimes referred to as IR ring nebulae~\cite{authors:92}, without any interpretation of this morphology. On the other hand, it was already proposed in \cite{2006ApJ...649..759C} that ring nebulae are actually projections of three-dimensional shells onto the plane of the sky, leading to the use of the alternate term ``bubble'', implying that these are three--dimensional objects.

Analyses of images of ring nebulae at 20~cm have shown that most of them are probably associated with regions of ionized hydrogen created by one or more O or B stars. In particular, Deharveng et al..~\cite{2010A&A...523A...6D} concluded that 86\% of the objects in the catalog~\cite{2006ApJ...649..759C} correspond to H II regions formed around massive, hot O--B2 stars.

The basis of the catalog \cite{2006ApJ...649..759C} is images at 8 $\mu$m. However, specific morphologies are also characteristic of IR ring nebulae at other IR wavelengths. Emission at 24 $\mu$m is observed inside nearly all rings or arcs of emission at 8 $\mu$m (see, e.g., \cite{2008ApJ...681.1341W}). It is usually believed that the 8 $\mu$m emission is mainly a manifestation of polycyclic aromatic hydrocarbons (PAHs) \cite{tielensaraa}. The absence of 8 $\mu$m emission or presence of only weak 8 $\mu$m emission inside ring nebulae could indicate that any PAHs inside the H II region have been partially or completely destroyed by UV radiation from the central star \cite{2013ARep...57..573P}. On the other hand, since the IR emission of PAHs is excited by the absorption of UV photons, the farther from the star the PAHs (or other small aromatic particles) are located, the weaker the intensity of their emission. Thus, the 8 $\mu$m ring apparently is located between the zone where PAHs are destroyed and where they are visible.

A less bright ring at 24 $\mu$m is also coincident with the outer ring of 8 $\mu$m emission, but most of the IR emission at 24 $\mu$m comes from the inner region. This appears as a central peak or region of extended emission, fairly often resembling a ring or arc with a smaller diameter. The emission at 70, 100, and 160 $\mu$m, which corresponds to larger dust grains, also appears as an outer ring surrounding the region of ionization. Here and below, we will refer to rings of emission with various diameters roughly coincident with the 8 $\mu$m ring of emission as outer rings, and the region inside the 8 $\mu$m ring of emission as the inner region.

Studies of the origins of the characteristic distributions of the IR emission in H II regions at various wavelengths are of considerable interest for our understanding of the evolution of various types of grains. As a first step toward such studies, Pavlyuchenkov et al. ~\cite{2013ARep...57..573P} analyzed dust emission using the MARION one--dimensional model for an expanding H II region~\cite{2009ARep...53..611K}. The dust was taken to be frozen in the gas, and the photodestruction of PAHs was taken into account using a phenomenological expression. It was shown that the absence of central 8 $\mu$m emission cannot be explained without including the photodestruction of PAHs. The appearance of an outer ring at 24 $\mu$m coincident with the 8 $\mu$m ring can be understood in this model as due to the stochastic heating of small carbon grains, but the presence of an inner ring at 24 $\mu$m remains unexplained.

Akimkin et al.~\cite{2015MNRAS.449..440A} used a modernized version of the MARION code to analyze the drift of charged dust due to radiation pressure in an H II region. This made it possible to qualitatively explain the inhomogeneous distribution of grains of various sizes inside H II regions, which can lead to an inner ring of emission at 24 $\mu$m.

A more detailed quantitative comparison of the results of simulations and observations requires a set of suitable real objects. The H II region RCW120 was treated as a ``reference'' object in ~\cite{2013ARep...57..573P, 2015MNRAS.449..440A}. It appears as a nearly perfect circle in IR images~\cite{2008hsf2.book..437D,2009A&A...496..177D}, suggesting that conclusions obtained in analyses of one--dimension models might be applicable to this object. In particular, the application of a one--dimensional dynamical model for the expansion of an H II region to RCW120 suggested the formation of young star clusters around this object, in a ``collect--and--collapse'' scenario \cite{2007A&A...472..835Z}.

The nearly perfect, regular shape of RCW120 attracted the attention of various other researchers, who also constructed dynamical models for this region. However, it is interesting that some studies have concluded that the observed characteristics of RCW120 require at least a two--dimensional dynamical model, for example, in which a young massive star is moving in a molecular cloud in this region, and the dust emission at 24 $\mu$m forms an arc around this star~\cite{2016A&A...586A.114M}). Torii et al.~\cite{2015ApJ...806....7T} investigated the idea that RCW120 formed during a collision of two gas--dust clouds, through which a massive young star is moving. In other words, this object, which appears at first glance to be a one--dimensional ``perfect bubble'', may prove to be very different from such a simple picture upon more detailed study, and far from all its properties can be described in a one--dimensional model.

On the other hand, the properties of the emission of grains with various sizes and the chemical compositions of H II regions are determined by a complex set of processes, not all of which can adequately be taken into account in available multi--dimensional hydrodynamical models. The resource intensive nature of the associated computations necessitates their implementation in only one dimension. Further, one--dimensional models can fully adequately describe the main properties of H II regions and their associated ring nebulae. As was noted above, the MARION onedimensional chemical--dynamical model is able to explain the key properties of IR ring nebulae~\cite{2013ARep...57..573P,2015MNRAS.449..440A}. One--dimensional dynamical models can also explain the ages of young star clusters, not only near RCW120, but also, for example, near W40~\cite{2013MNRAS.436.3186P} and S235~\cite{2014MNRAS.437.1593K}. It is possible that onedimensional models can also explain other properties of the IR emission of H II regions, such as so--called ``yellowballs''~\cite{2015ApJ...799..153K}.

Comparison with a single spherically symmetrical object is not sufficient if we wish to trace the entire evolutionary path of an H II region from its compact to its developed stage using chemical--dynamical simulations together with radiative--transfer computations and the construction of theoretical IR maps. A more extensive set of objects is needed for comparisons of the results of evolutionary simulations with the results of IR observations obtained with the Spitzer, Herschel, and other telescopes for specific stages in the development of H II regions. Our aim in this study was to analyze the morphologies of H II regions manifest in IR images as ring nebulae, and to distinguish a set of these objects with close to circular shapes, making them the most promising objects for one--dimensional modeling.

Section 2 describes the data used and the methods applied to reduce them. Section 3 is dedicated to an analysis of the morphologies of IR images of nebulae. We discuss the results obtained in Section 4, and formulate our conclusions in Section 5

\section{DATA AND THEIR REDUCTION}\label{sec:data}

\subsection{Sources of Data}\label{sec:datacat}
The 20-cm New GPS survey\footnote{http://third.ucllnl.org/gps/}, created using the MAGPIS~\cite{Helfand_MAGPIS} database of radio images of regions with Galactic coordinates $|b_{\rm gal}| < 0.8^{\circ}$ и $5^{\circ} < l_{\rm gal} < 48.5^{\circ}$, was used as the basis for this study. We identified sources of compact radio emission among the objects in this survey, toward which we carried out a visual search for ring nebulae at 8 and 24 $\mu$m using images obtained with the IRAC~\cite{2004ApJS..154...10F} and MIPS~\cite{2004ApJS..154...25R} cameras of the Spitzer Telescope. The resulting list contains objects having the apperance of rings at 8 $\mu$m, inside of which IR emission at 24 $\mu$m and radio emission at 20 cm is observed. Moreover, assuming that the part of the spherically symmetrical shell that is facing the observer should be optically thick in the visible, we also checked for an absence of H$\alpha$ when possible, although we were not able to find such data for all the objects. The total number of selected objects was 99, for which we also analyzed 70 $\mu$m images from the Herschel Telescope archive obtained with the PACS instrument \cite{PACS}. Six objects in our sample (S15, S21, S44, S123, S145, and S167) did not fall into the 20 cm New GPS survey, but were identified earlier in  \cite{2006ApJ...649..759C} as IR ring nebulae associated with regions of ionized hydrogen. We decided to include them in our study, sice they are morphologically very similar to the remaining objects.

\subsection{Cleaning the Images of Point Sources}\label{sec:datapoint}

Prior to our analysis of the morphology of the IR images of ring nebulae and the construction of radial-intensity profiles, we cleaned the archival IR maps used of point sources, since we are interested in the distribution of emission in these objects. We developed Python code to carry out this cleaning in an automated regime. The search for and removal of the point sources was conducted in three stages, corresponding to the removal of images of bright stars, stars of moderate brightness, and faint stars. Modules in the code were called three times in this process, with appropriate parameters.

The specified parameters were the saturation level for the image (10 000 in digitcal readouts, ADU), the lower intensity threshold for the detection of a point ource (500, 50, or 5 pixels for bright, intermediate, and faint sources, respectively), a rough estimate of the dimensions of the image of a point source for the indicated threshold intensity (FWHM = 5, 2, and 2 pixels, respectively), and the lower intensity threshold for the detection of extended objects (100, 10, 1 pixels, respectively).

Most importantly, the data were cleaned of high--frequency noise at each stage~\cite{1965ApJ...142..591O}. This was done using a convolution with the kernel
$$
K_{3\times3}= 
\left(
\begin{array}{ccc}
1/4 & 1/2 & 1/4 \\
1/2 & 1 & 1/2 \\
1/4 & 1/2 & 1/4 \end{array}
\right).
$$

\begin{figure}[h]
\centering
\includegraphics[width=0.45\linewidth]{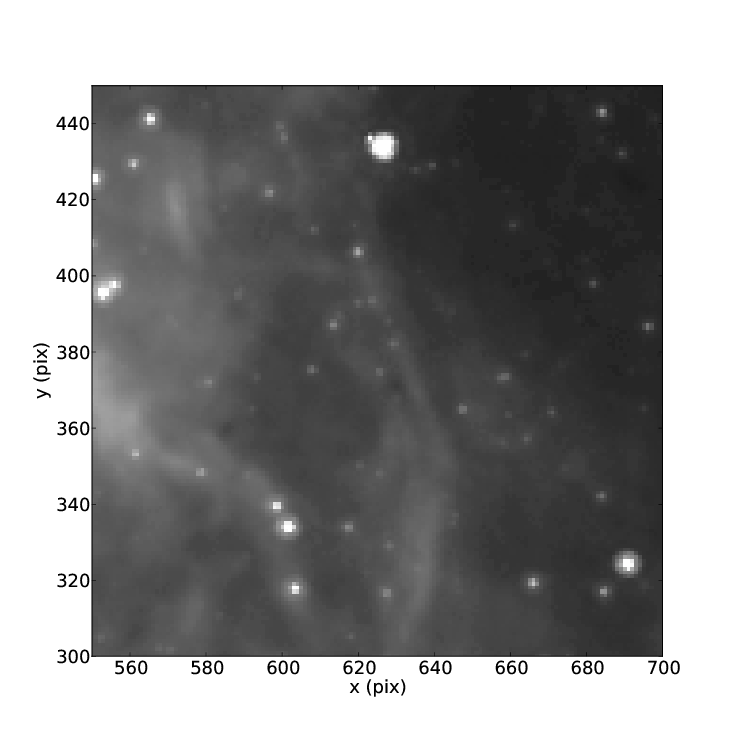}
\includegraphics[width=0.45\linewidth]{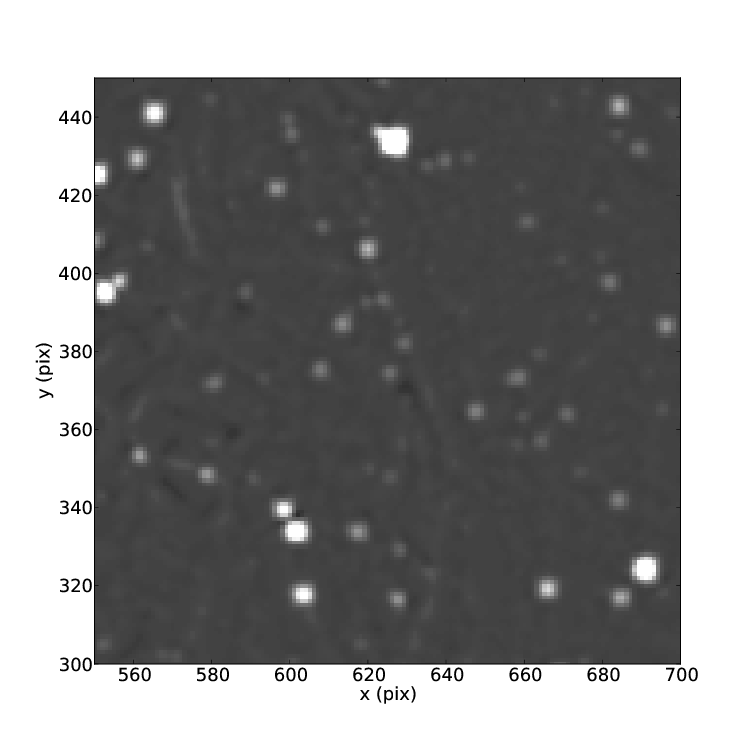}
\caption{Original image (left) and filtered image with the local background removed used to subtract point sources (right).}\label{fig:1_2}
\end{figure}

Further, a median filter with core size $3\times3$ pixels was used to estimate the local background levels in the image. The resulting background values were subtracted from the image. This dataset was also passed through a low--frequency filter. Figure~\ref{fig:1_2} presents an example of an original image and the filtered image used to search for the point sources.

\begin{figure}[t!]
\center{\includegraphics[clip=,width=0.5\linewidth]{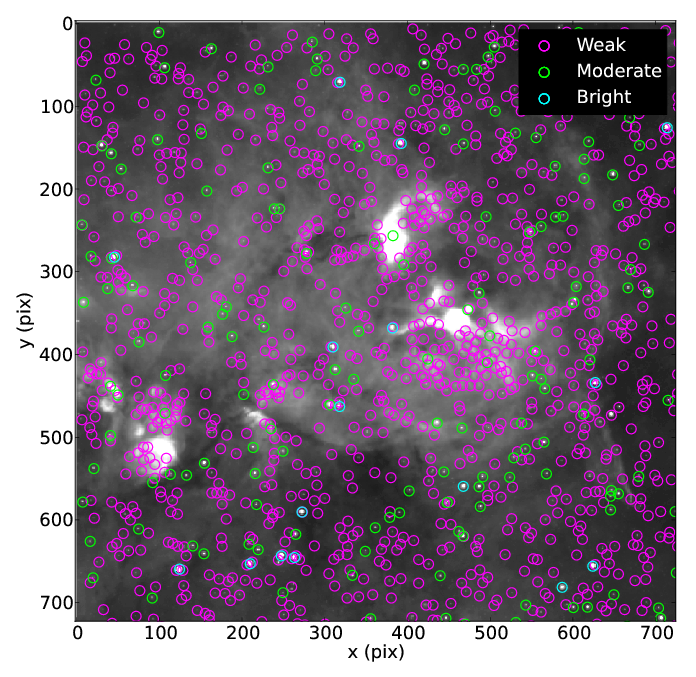}}
\caption{Example of identifying point sources. The different colors correspond to objects with different brightnesses.}
\label{ris:3}
\end{figure}

We further identified regions containing pixels with values higher than the threshold intensity for the point and extended objects. If the size of the region exceeded 100 pixels, it was taken to be extended. The results of our search for point objects are shown in Fig.~\ref{ris:3}.

We delineated a region of interest for each point source found (FWHM~$\times 4$), using the position of the pixel with the maximum intensity as an approximate estimate of the position of the center of the source. We refined the source positions via least-squares fits of the profiles with Gaussians. We then subtracted the model Point Spread Function (PSF) from the image. The amplitude of the PSF was determined from the least-squares fit, together with the center of the image. The model PSF of the Spitzer telescope was obtained from an analysis of the mean PSFs for different sections of the field of view (Fig. 3). The three--beam symmetry of the first diffraction ring of the PSF was not taken into account. The residual intensity after subtraction of the model PSF comprises about 4\% of the intensity of the central maximum.

\begin{figure}[t!]  
\includegraphics[height=0.3\textwidth]{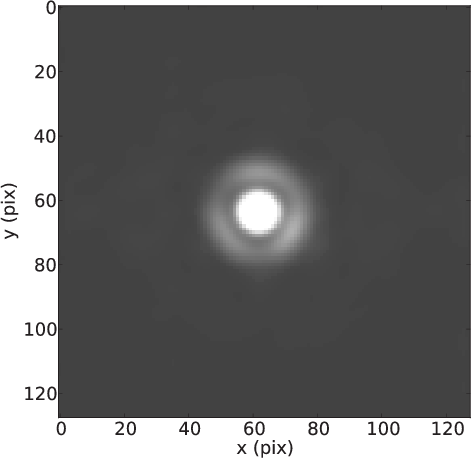}
\includegraphics[height=0.3\textwidth]{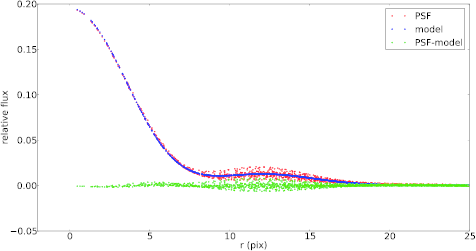}
\caption{Upper: Field--averaged PSF of the Spitzer telescope. Lower: Result of modeling the PSF; the real PSF is shown in red, the model PSF in blue, and the residual intensity in green..}
\label{fig:4_5}
\end{figure}

\subsection{Morphology of the IR Images of the Nebulae}\label{sec:ellipse}

We obtained an approximate description of the morphologies of the IR ring nebulae by fitting their images with ellipses. This enabled determination of the center of the nebula, as well as its size, degree of asymmetry, and orientation to the plane of the sky. The fitting of the IR images with ellipses was conducted after point sources were removed (at 8 and 24 $\mu$m), and was carried out in three stages.

\begin{figure}[h]
\includegraphics[width=0.5\linewidth]{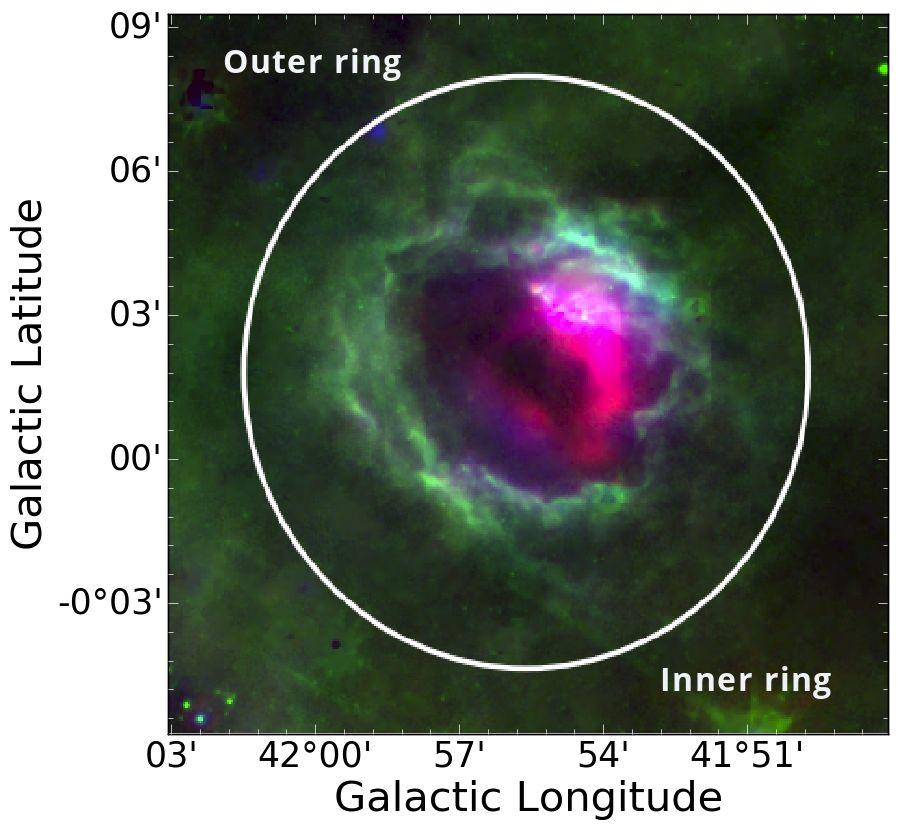}
\caption{Example of a region used to determine the photometric center of the object N49 from [2]. The emission at 8 $\mu$m (outering) is shown in green, the emission at 24 $\mu$m (inner ring) in red, and the emission at 70 $\mu$m.}
\label{ris:region_circle}
\end{figure}

\begin{enumerate}
\item We first found the photometric center of the image ($x_0$, $y_0$) by determining the ``center of mass'' of the intensities in all the pixels in a selected region around the object (Fig.~\ref{ris:region_circle}).

\item We then constructed rays extending in all directions from the photometric center, on which we identified the coordinates of pixels with the maximum intensities.

\item The resulting contour around the photometric center was approximated with an ellipse via a leastsquares fit. This yielded the following parameters of the region: the geometric center (center of the fitted ellipse), major and minor axes ($a$, $b$), and position angle $\phi$ relative to the direction toward the Galactic North pole.
\end{enumerate}

\begin{figure}[t!]
\includegraphics[width=0.3\textwidth]{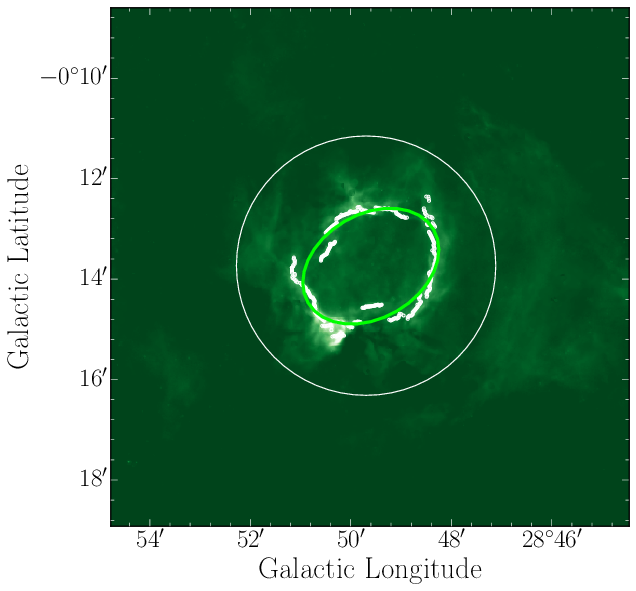}
\includegraphics[width=0.3\textwidth]{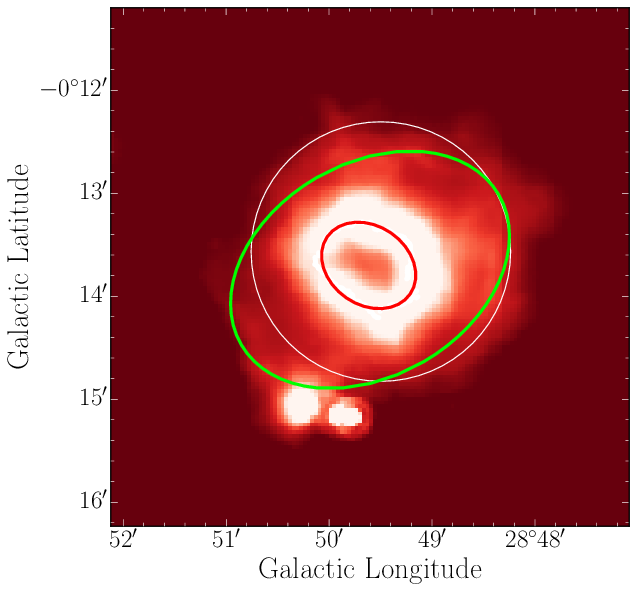}
\includegraphics[width=0.3\textwidth]{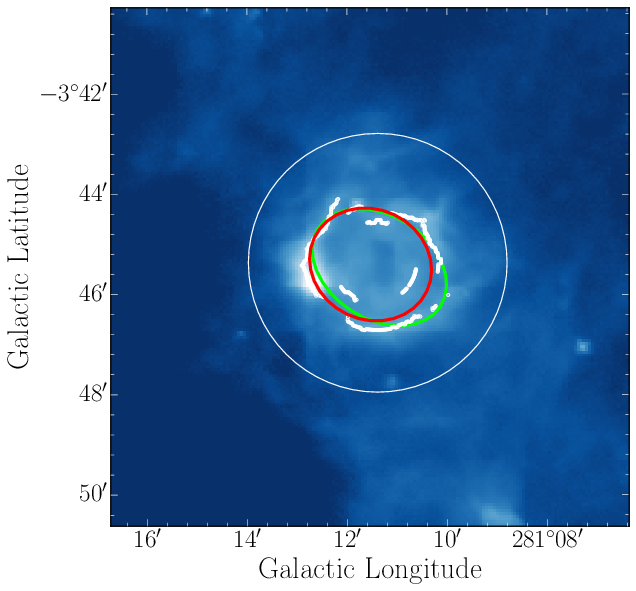}
\caption{Examples of ellipses fitted to the contours of the object N49 at 8 $\mu$m (left), 24 $\mu$m (center), and 70 $\mu$m (right). The ellipse fitted to the 8 $\mu$m contour is shown in green in the center and right panels. The white circles show the regions used in our analysis.}
\label{fig05}
\end{figure}

\begin{figure}[h]
\center{\includegraphics[width=0.5\linewidth]{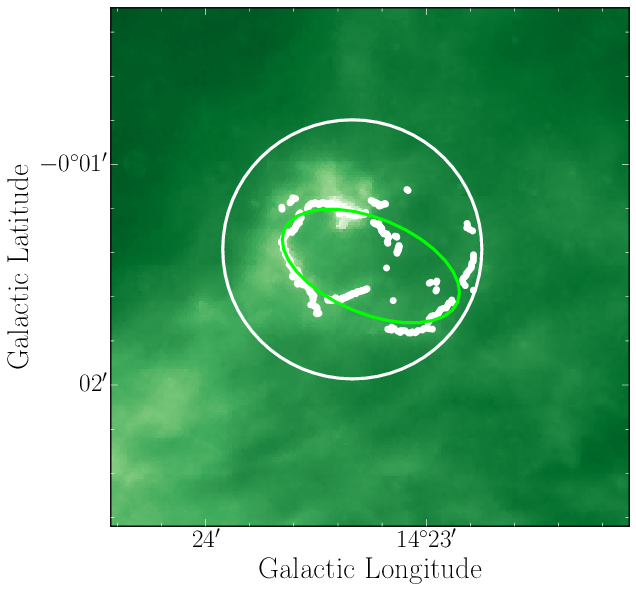}}
\caption{Example of an ellipse that poorly fits the 8 $\mu$m ring emission of the object MWP1G014390--000200S}
\label{ris:2111}
\end{figure}

The procedure for fitting the ellipses was carried out for the maps at 8, 24, and 70 $\mu$m. The resolution of the longer-wavelength maps was insufficient for this analysis. Examples of images of ring nebulae with fitted ellipses are shown in Fig.~\ref{fig05}. Note that it was not possible to obtain a correct fit with an ellipse in a number of cases, due to poor resolution and/or complex structure of the object. An example is shown in Fig.~\ref{ris:2111}. The formal errors in the semiaxes of the ellipses are about $0.1''$, but, taking into account the uncertainty in the fitting, we estimate the uncertainty in the size of the object to be roughly $1''$(with a corresponding uncertainty in the eccentricity of $\sim0.1$).

The procedure used to choose an elliptical contour was carried out such that the ellipse was fitted in the outer ring at 8 and 70 $\mu$m and in the inner ring at 24 $\mu$m (examples of the arrangement of the rings are shown in Fig.~\ref{ris:region_circle}). The inner ring can also be visually distinguished on images at wavelengths from 70 to 160 $\mu$m, but, for most objects, the corresponding intensity and/or angular resolution is insufficient for the automated analysis. Thus, the innerring morphology must be studied individually at these wavelengths. The fitting enables the construction of the eccentricity and position angle distributions for the objects, and also checks for possible correlations between these parameters. The results are discussed in the following section.

\section{RESULTS OF THE ANALYSIS}\label{sec:results}

\begin{figure}[t!]
\includegraphics[width=0.4\textwidth]{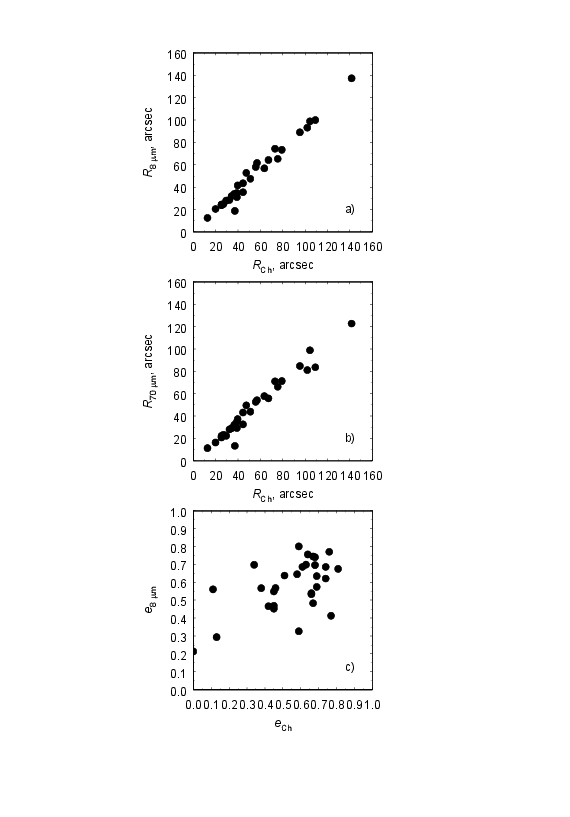}
\caption{Comparison of the computational results from our study (vertical axis) with the results from the catalog of Churchwell et al.~\cite{2006ApJ...649..759C} (horizontal axis, subscript ``Ch''). Shown are comparisons of the sizes of the nebulae estimated from (a) 8 and (b) 70 $\mu$m images and the (c) eccentricities of the nebulae estimated from the 8 $\mu$m images.}
\label{churchcompar}
\end{figure}

The results of fitting ellipses to the 8 and 70 $\mu$m images of the nebulae are presented in Tables \ref{tab:catal8mkm} and \ref{tab:catal70mkm}. A dash indicates that fitting is not possible for that object. Of our selected objects, 86 are considered earlier in~\cite{2006ApJ...649..759C,2012ApJ...759...96B, 2012MNRAS.424.2442S}. We also identify four new objects in the studied regions of the sky, denoted TWKK in the tables. The sizes and eccentricities of 30 objects determined in~\cite{2006ApJ...649..759C} are compared with our results in Fig.~\ref{churchcompar}.

The sizes of the shells at both 8 and 70 $\mu$m are in good agreement with previous estimates (we adopted $R=\sqrt{ab}$ as an estimate of the size). The situation with the eccentricities is less clear: the values of $e$ in our study and in ~\cite{2006ApJ...649..759C} are correlated, but with substantial scatter, possibly due to uncertainty introduced when fitting the ellipses to the images in order to estimate their parameters in~\cite{2006ApJ...649..759C}, which is done by hand.

\begin{figure}[t!]
\includegraphics[width=0.5\textwidth]{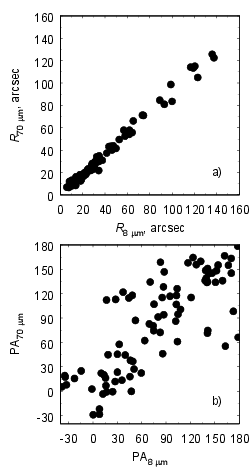}
\caption{Comparison of the (a) sizes and (b) position angles of the ellipses fitted at 8 and 70 $\mu$m.}
\label{PA}
\end{figure}

Figure~\ref{PA} compares the sizes and position angles of the fitted ellipses at 8 and 70 $\mu$m. Both parameters are in good agreement with each other, suggesting a single physical nature for the shells radiating at these two wavelengths. The position-angle distributions of the ellipses at 8 and 70 $\mu$m are essentially uniform, indicating an absence of a preferred orientation for the ring nebulae (such as would be expected in the presence of large-scale structures in the studied regions of sky). The mean relative difference in the positions of the ellipse centers is approximately $(0.1-0.2)R_8$, where $R_8$ is the size of the object at 8 $\mu$m. As a rule, larger values are characteristic for objects with complex morphologies or compact objects whose structures are poorly defined at 70 $\mu$m.

\begin{figure}[t!]
\includegraphics[width=0.5\textwidth]{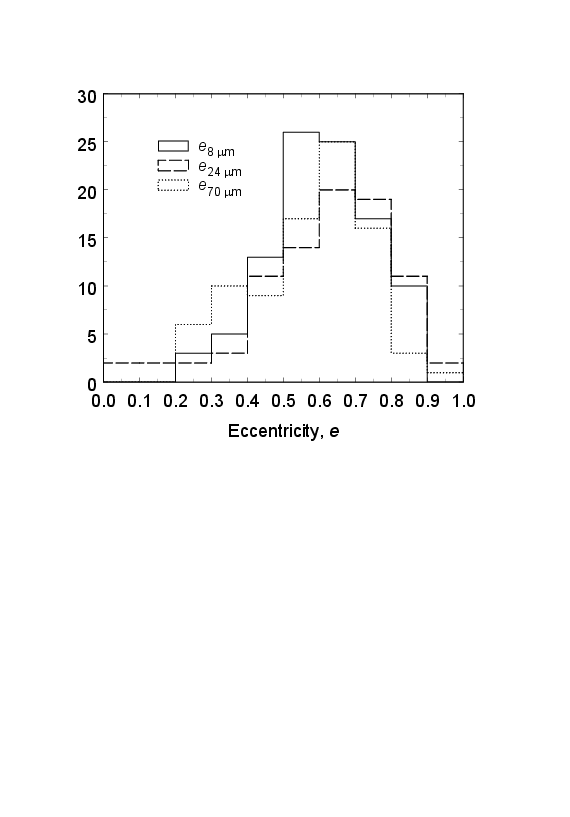}
\caption{Eccentricity distribution of the objects.}
\label{ris:excent}
\end{figure}

The distributions of the eccentricities of the ellipses fitted to the objects at 8, 24, and 70 $\mu$m are shown in Fig.~\ref{ris:excent}. The e values for roughly half the objects do not exceed 0.6 (this is the median eccentricity at 8 and 24 $\mu$m); given the uncertainty in distinguishing the shells, this suggests they have close to circular shapes. Note that the highest eccentricities are indicated not by an elongated shape of the shell, but instead by its irregular structure (e.g., not being closed): high eccentricities are accompanied by appreciable discrepancies in the positions of the ellipse centers at 8 and 70 $\mu$m, underscoring the complexity of the structures of these objects.

The results of fitting ellipses to the 24 $\mu$m maps deserve a separate discussion. As was noted above, the fitting procedure was carried out so that the 24 $\mu$m ellipses traced the inner emission (relative to the shell at $\mu$m). Strong non-uniformity of this emission means that the parameters of the ellipses reflect their real structure only in some cases (as, for example, in the middle panel of Fig.~\ref{fig05})). Figure~\ref{ris:excent} shows that some of the ellipses fitted at 24 $\mu$m have eccentricities lower than 0.2. All of these ellipses are very small in size (compared to $R_8$, and their centers are appreciably shifted relative to the centers of the ellipses at 8 and 70 $\mu$m. This result apparently corresponds to the situation when a very bright peak is distinguished in the inner ring, while the remainder of the ring is lost against the overall background. Ellipses with high eccentricities whose centers nearly coincide with the centers of the outer ellipses correspond to unclosed inner rings. Even when the inner ring is closed, its orientation can differ stronglly from the orientation of the outer ring. For example, the position angles of the 8 and 24 $\mu$m ellipses in the object N49 differ by 115$^\circ$, although both rings are practically perfectly closed.

\begin{figure}[h!]
\includegraphics[width=0.3\textwidth]{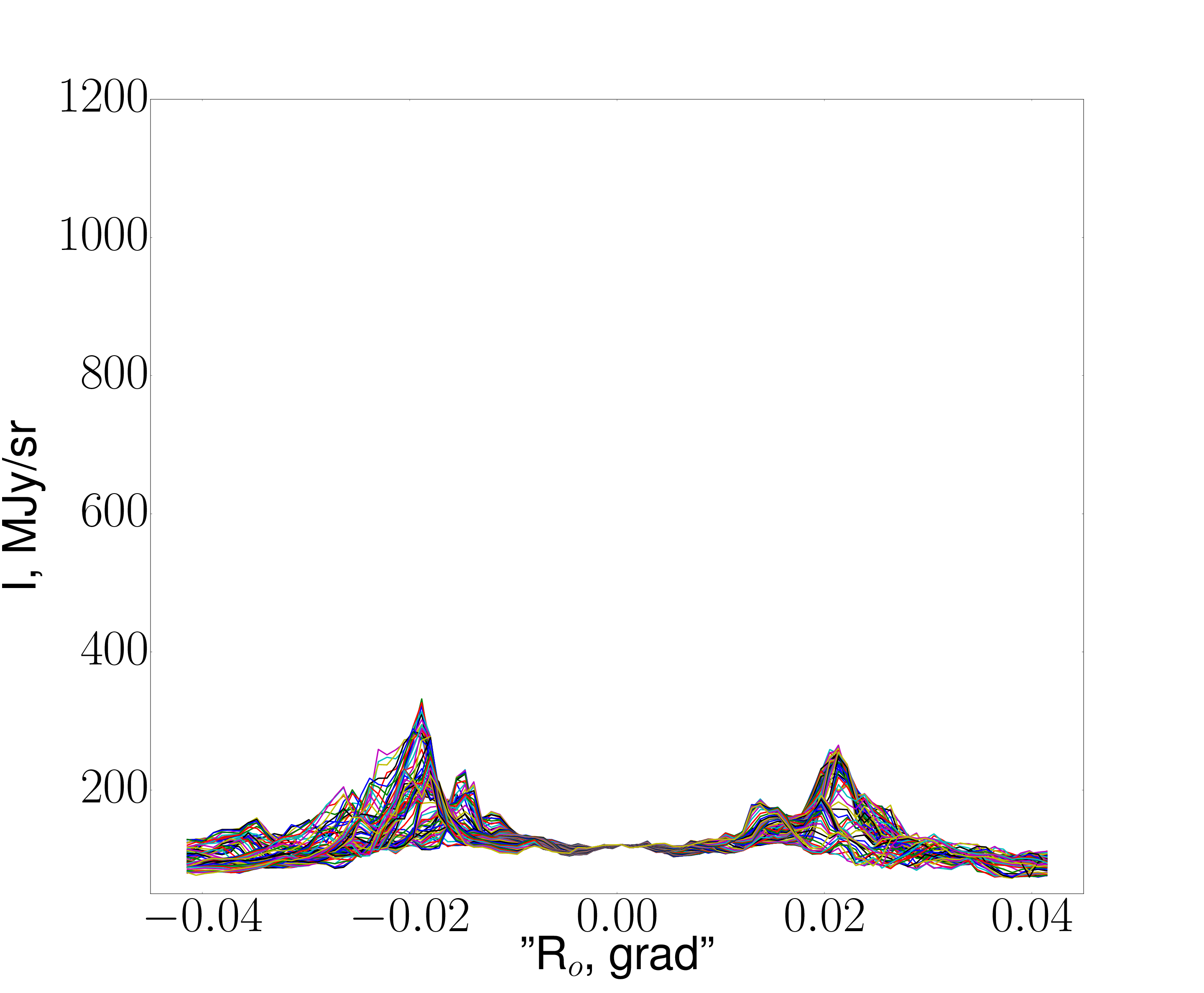}
\includegraphics[width=0.3\textwidth]{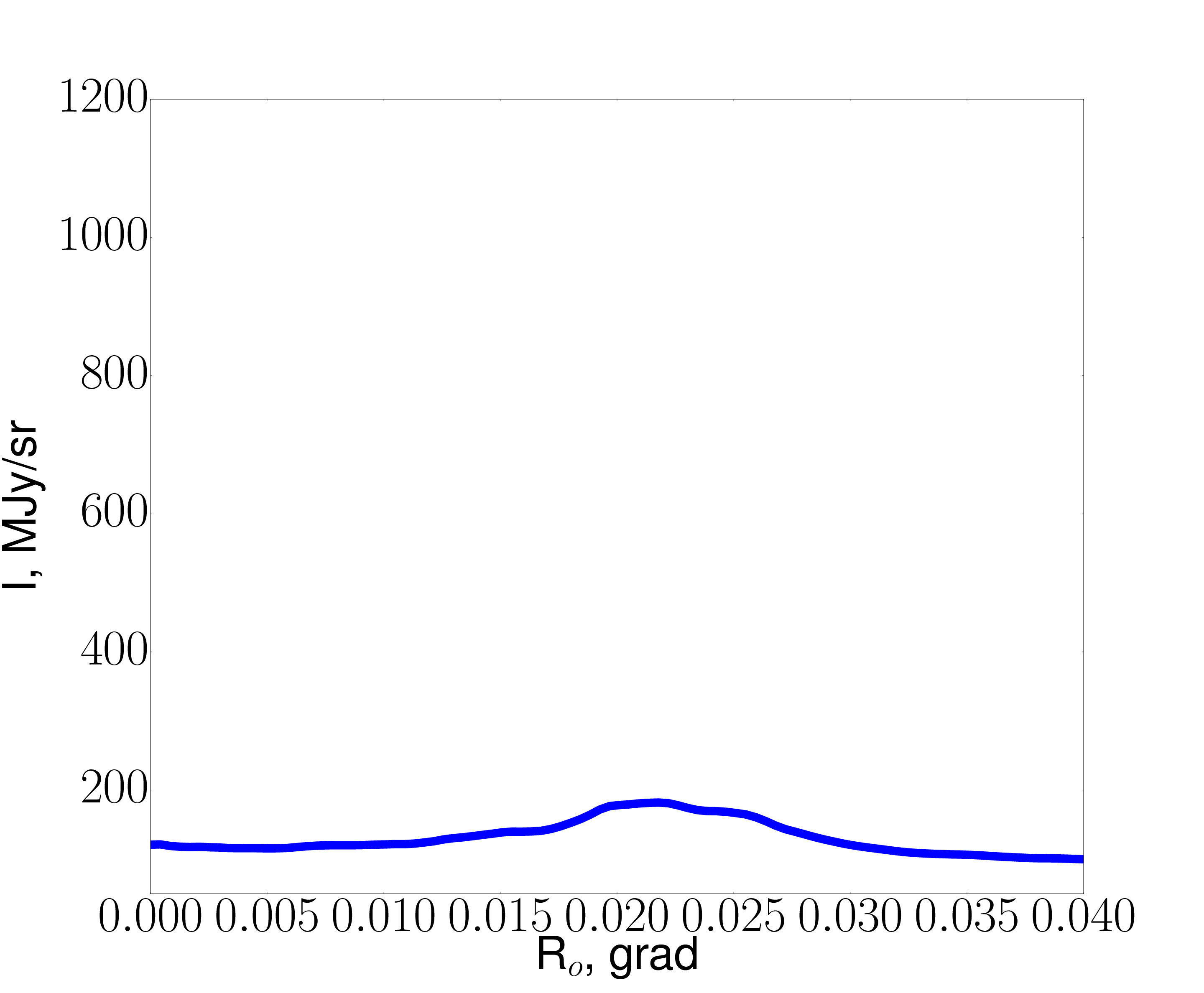}\\
\includegraphics[width=0.3\textwidth]{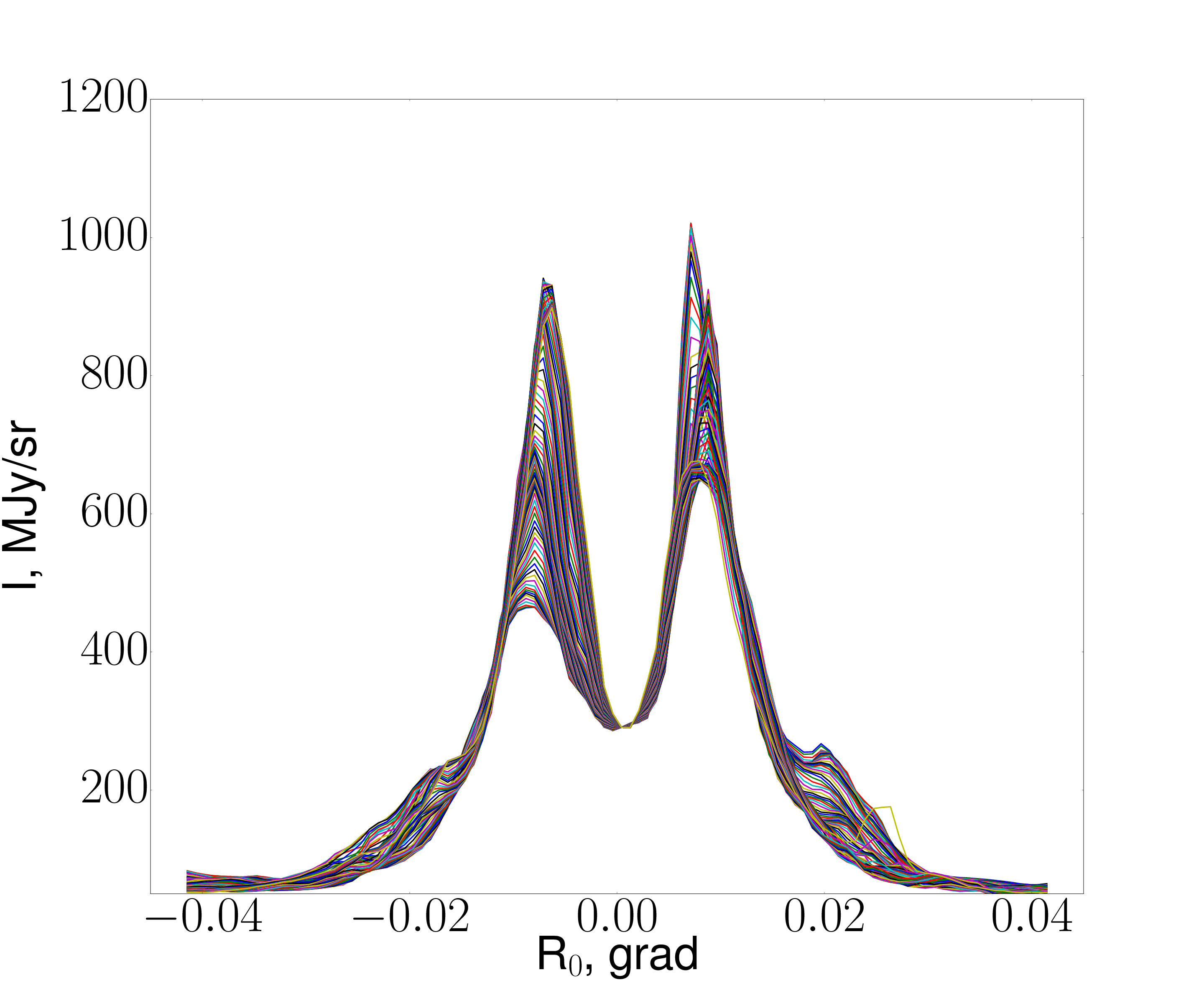}
\includegraphics[width=0.3\textwidth]{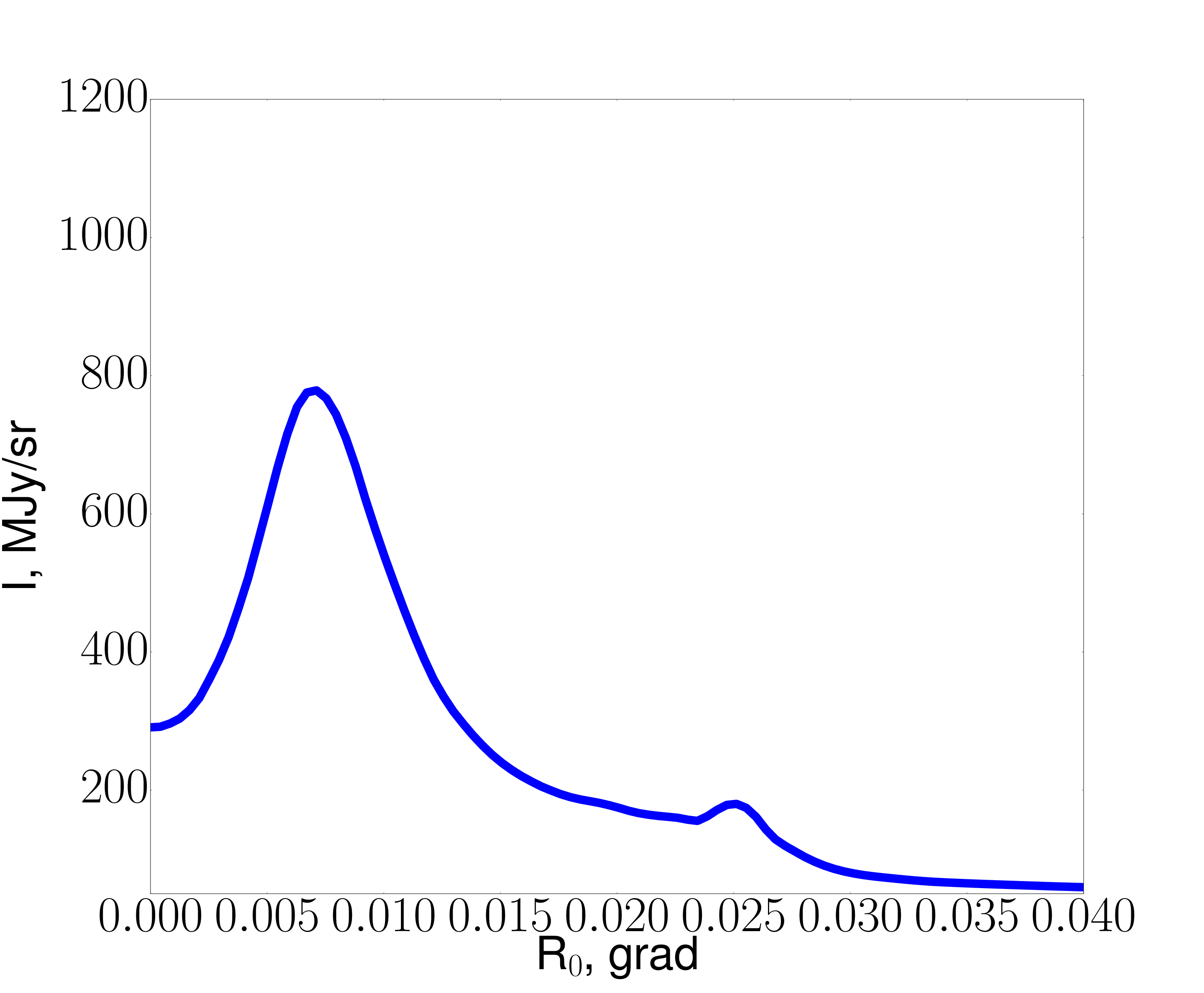} \\
\includegraphics[width=0.3\textwidth]{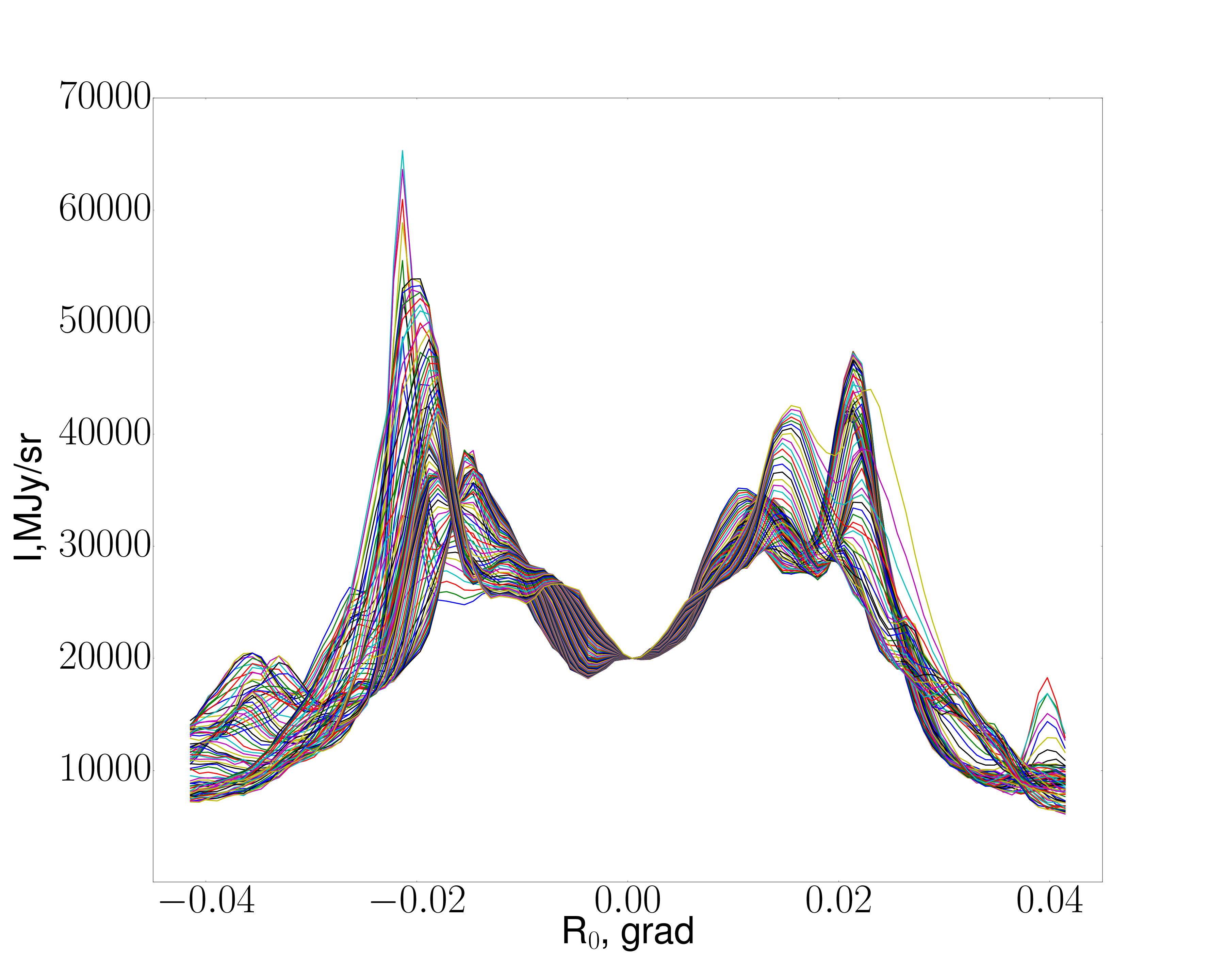}
\includegraphics[width=0.3\textwidth]{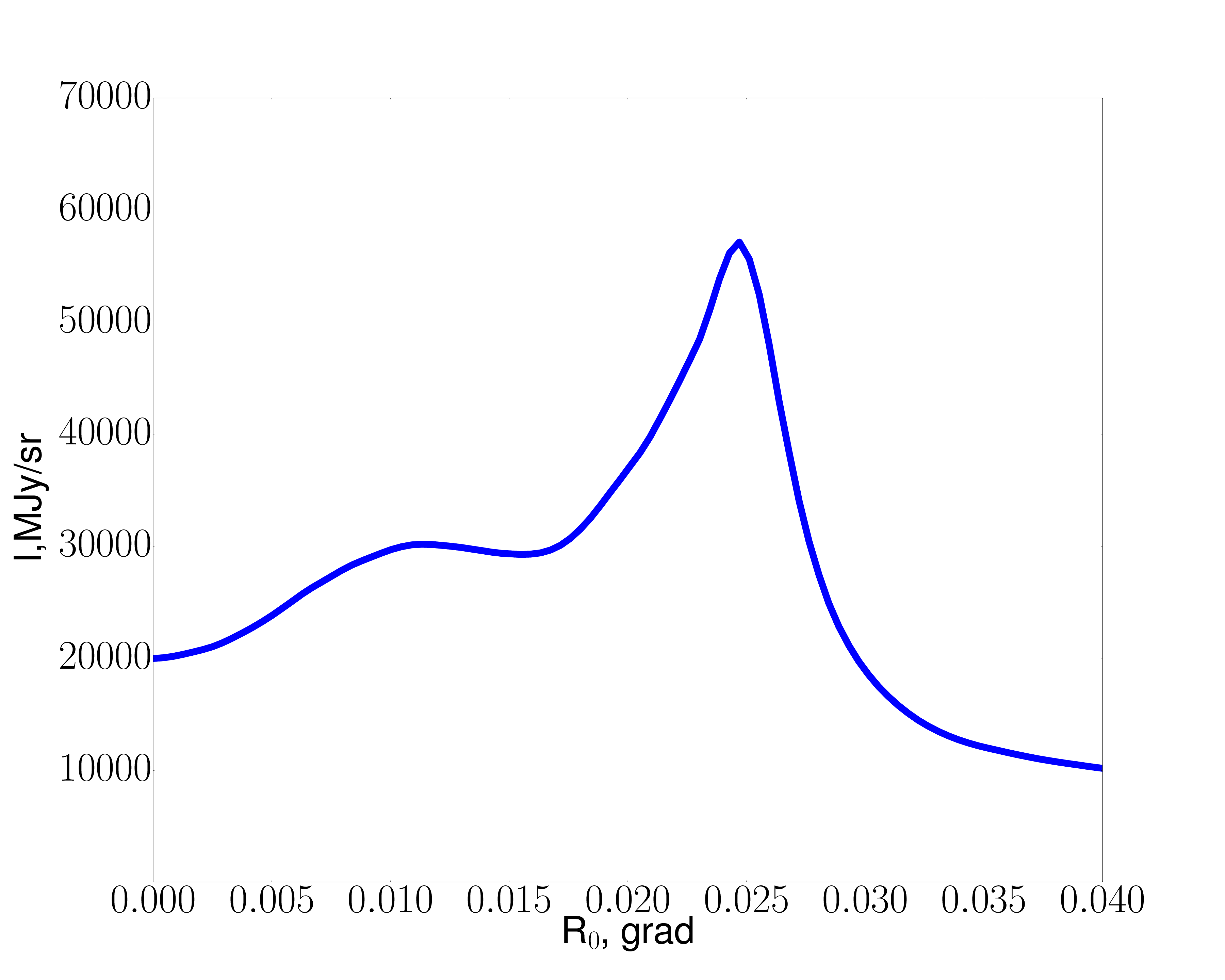}
\caption{Radial--intensity profiles for the object N49 at 8 $\mu$m (upper row), 24 $\mu$m (middle row), and 70 $\mu$m (lower row). A set of intensity profiles for 256 directions (left) and an azimuth-averaged intensity profile (right) are shown. The vertical axis plots the intensity in MJy/ster, and the horizontal axis distance from the derived geometrical center.}
\label{ris:N49_all_sred}
\end{figure}

In addition to fitting the images with ellipses, we also derived radial--intensity profiles of the nebulae at 8 and 24 $\mu$m, constructing 256 rays through the center of the ellipse for each object (a detailed analysis of the photometric properties of the studied objects will be presented in a separate study). Examples of profiles for the object N49 shown earlier in Fig. \ref{fig05} are presented in Fig.~\ref{ris:N49_all_sred}, where the profiles for the different position angles are denoted by different color curves. Figure~\ref{ris:N49_all_sred} also shows the radial profiles averaged in azimuth, which can be used to compare with the results of numerical simulations.

The intensity profiles at 8 and 24 $\mu$m are presented on the same vertical scale in order to emphasize the appreciable dominance of the emission of the inner ring at 24 $\mu$m over the emission of the outer ring at both 8 and 24 $\mu$m. The outer rings in N49 have nearly the same mean size at the three wavelengths; in other objects, they can be shifted slightly relative to each other. It is interesting to note that the weak inner ring emission at 70 $\mu$m is located slightly farther from the center of the object than the ring at 24 $\mu$m. Such inner rings at long wavelengths ($\lambda\ge70$ $\mu$m) are also observed in other objects. In the small number of cases when the morphology of an object is fairly regular and well resolved, we have the impression that the ring of emission at 24 $\mu$m goes along the inner edge of the ring of emission at 70 $\mu$m. We have not analyzed the detailed structure of the ring nebulae at long IR wavelengths here, but plan to examine this in detail in a future study dedicated to an analysis of the photometric properties of the studied objects.

A comparison of the sizes of the ellipses ($\sqrt{ab}$) at 8 and 24 $\mu$m shows that the mean ratio $R_{8}/R_{24}$ is about 2.5 for our objects. The maximum value of this ratio exceeds 40, which in reality corresponds to a peak of the 24 $\mu$m emission rather than a ring. This agrees with the results of theortical simulations of the IR emission from H II regions~\cite{2013ARep...57..573P}, which indicate that PAHs radiate 8 $\mu$m emission from the photodissociation region around the H II region, while the emission at 24 $\mu$m is produced by larger carbon grains that are able to survive inside the H II region. There are also several objects whose ellipses at 24 $\mu$m are only slightly smaller than their ellipses at 8 $\mu$m; this corresponds to the situation when the ring at 24 $\mu$m is directly adjacent to the inner boundary of the ring at 8 $\mu$m. In all cases, the maximum intensity of the inner emission at 24 $\mu$m appreciably exceeds the maximum intensity of the outer ring at 8 $\mu$m. The origin of the differences in the sizes and the relationships between them is a question that can be addressed through one--dimensional modeling. It is possible that the different relative sizes of the rings (central peaks) at 24 $\mu$m in different objects reflect their different ages.

\section{DISCUSSION}

The term ``infrared bubble'' is widely used in the literature for ring nebulae, implying that these nebulae are the projections of structures having {\em roughly\/} spherically symmetric shapes. However, discussions about whether some of them are indeed rings, i.e., two-dimensional structures, have continued. Such rings or tori could arise around hot stars located inside relatively thin, dense molecular layers \cite{sheet}. One objection to this hypothesis is that, if ring nebulae is formed in thin layers, we should observe them not only ``face on'', but also ``side on'', in the form of bipolar nebulae forming as a result of the gaps in expanding H II regions on either side of the layer. Such objects are indeed observed \cite{bip}, but their number is not sufficient to elucidate whether they are widespread and related to ring nebulae.

Statistical studies of the shapes of ring nebulae can shed light on their nature. Such studies were carried out in \cite{2016PASJ...68...37H}, where IR nebulae were analyzed using observations obtained on the Spitzer and AKARI telescopes at wavelengths from 8 to 160 $\mu$m. Hattori et al. \cite{2016PASJ...68...37H} classified these objects by type (closed ring, open ring, unclassified object), and also investigated the relationships between their radii, luminosities, and relative luminosities at various wavelengths. They showed that the IR nebulae of all morphologies have some common properties, with the exception of large regions with open outer rings and high total IR luminosities. Hattori et al. \cite{2016PASJ...68...37H} suggest that the origin of such nebulae could be related to collisions of two gas--dust clouds.

Similar explanations have been proposed for the nebula RCW49 \cite{2009ApJ...696L.115F}, and even the ``perfect bubble'' RCW120 \cite{2015ApJ...806....7T}. Evidence for a possible collision in the past includes the presence of molecular clouds with various radial velocities associated with the IR nebula. The large difference in the velocities of two (about 15 km/s) clouds indicates that they cannot be gravitationally bound. These facts testify that both the formation of massive ionizing stars in these regions and the structure of the matter around them may have initiated a collision of two clouds. The example of RCW120 shows that a model with colliding clouds could explain the kinematics and morphology of the gas not only in regularly shaped ring nebulae, but also in irregular nebulae with open shapes.

The main aim of creating our sample of objects was to select ring nebulae for comparison with the results of one-dimensional numerical simulations. In other words, we wished to identify objects whose shapes were not too elongated (we adopted a limiting eccentricity of $e=0.6$ for the fitted ellipses at 8 $\mu$m), resembling shells, not rings. One type of evidence for shell structure for at least some objects is the absence of line emission, in particular H$\alpha$. The numerical simulations conducted in \cite{2013ARep...57..573P} show that the optical depth of the leading wall of a spherically symmetrical shell of ionized hydrogen should be at least a few tens in the visible. We found 47 such objects in our sample. However, in some of these, their low eccentricities reflect insufficient angular resolution (even at 8 $\mu$m) to determine their structure, rather than a symmetrical shape. If we introduce the additional constraint $R_8>20''$ (ten times the angular resolution at 8 $\mu$m), 32 candidate objects remain, which is sufficient for a statistical comparison of the results of observations and these theoretical computations.

Half the 8 $\mu$m fitted ellipses for the objects have eccentricities exceeding 0.6, testifying that the object is either very elongated or its outer shell is appreciably disconnected (i.e., the ellipse was essentially fit to some segment of the shell). Such distorted shells could arise during the expansion of a region of ionized hydrogen in a relatively thin molecular layer or in a strongly inhomogeneous cloud, and also in a collision of clouds. In all such cases, it is not possible to judge the nature of such a shell without additional detailed observations together with three-dimensional numerical simulations.

\section{CONCLUSION}\label{conc}

We have compiled a list of IR ring nebulae observed in the region $5^{\circ} < l_{\rm gal} < 48.5^{\circ}$ и $|b_{\rm gal}| < 0.8^{\circ}$. This was done using IR images of nebulae at 8, 24, and 70 $\mu$m, as well as radio images at 20 cm. The catalog contains 99 objects, primarily regions of ionized hydrogen around young, massive stars or small groups of such stars. To investigate the morphology of the observed rings, we fits ellipses to maps of the emission at 8, 24, and 70 $\mu$m. Among the sample objects, 32 has eccentricities for their 8 $\mu$m fitted ellipses not exceeding 0.6 and angular radii exceeding $20''$, indicating that their structure is resolved at 8 $\mu$m. The radial profiles of their IR emission are suitable for use as test distributions in one-dimensional modeling of these objects. The position-angle distribution of the fitted ellipses are roughly uniform, consistent with an absence of large-scale structure in the studied region. The sizes of the outer rings of emission at 8 and 70 $\mu$m are in good agreement with each other. The morphology of the inner emission at 24 $\mu$m (relative to the shell at 8 $\mu$m) is more complex, and can be described as an inner ring or peak of the emission only in some cases. The centers of the emission at 8 and 24 $\mu$m are sometimes appreciably non-coincident, suggesting strong inhomogeneity of the emission at 24 $\mu$m.

\section{ACKNOWLEDGMENTS}

This work was supported by the Russian Foundation for Basic Research (grants 16-02-00834 and NSh-9951.2016.2). This work has made use of the Astropy package \cite{Astropy}.


\newpage
{\scriptsize
\begin{longtable}{|p{5cm}|p{2cm}|p{2cm}|p{1cm}|p{1cm}|p{1cm}|p{1cm}|}
\caption{Results of elliptical fits for the shapes of objects in 8 $\mu$mm images. Objects were taken from: 1)Becker et al.~\cite{1994ApJS...91..347B}, 2) Simpson et al.~\cite{2012MNRAS.424.2442S}, 3) Churchwell et al.~\cite{2006ApJ...649..759C}, 4) new objects, 5) Urquhart et al.~\cite{2009A&A...507..795U}, 6) Egan et al.~\cite{2003yCat.5114....0E}.}\label{tab:catal8mkm}\\ \hline
Object & $l_{\rm gal}$, $^{\circ}$ & $b_{\rm gal}$, $^{\circ}$ & $a$, $^{''}$ & $b$, $^{''}$ & $e$ & PA, $^{\circ}$\\ \hline
\endfirsthead \hline
\multicolumn{7}{|c|}{\scriptsize\slshape(to be continued)} \\ \hline
Object & $l_{\rm gal}$, $^{\circ}$ & $b_{\rm gal}$, $^{\circ}$ & $a$, $^{''}$ & $b$, $^{''}$ & $e$ & PA, $^{\circ}$\\ \hline
\endhead \hline
\multicolumn{7}{|c|}{\scriptsize\slshape (to be continued)} \\ \hline
\endfoot \hline
\endlastfoot
S15$^{3}$ & 343.916 & --0.648                    & 140 & 108 & 0.64 & 12\\
S21$^{3}$ & 341.358 & --0.288                    & 51  & 37  & 0.69 & 178\\
S44$^{3}$ & 334.524 & 0.820                      & 156 & 121 & 0.64 & 87\\
S123$^{3}$ & 312.978 & --0.433                   & 149 & 124 & 0.56 & 133\\
S145$^{3}$ & 308.717 & 0.623                     & 382 & 272 & 0.70 & 139\\
S167$^{3}$ & 301.627 & --0.345                   & 389 & 364 & 0.35 & 95\\
CN67$^{3}$ & 5.526 & 0.037                       & 54  & 32  & 0.80 & 33\\
CN77$^{3}$ & 6.139 & --0.640                     & 67  & 48  & 0.70 & 75\\
CN79$^{3}$ & 6.202 & --0.334                     & 70  & 54  & 0.64 & 121\\
CN111$^{3}$ & 8.311 & --0.086                    & 99  & 88  & 0.47 & 37\\
MWP1G008430--002800S$^{2}$ & 8.431 & --0.276     & 21  & 9   & 0.89 & 40\\
CN116$^{3}$ & 8.476 & --0.277                    & 22  & 16  & 0.69 & 75\\
N4$^{3}$ & 11.893 & 0.747                        & 130 & 112 & 0.51 & 140\\
MWP1G012590--000900S$^{2}$ & 12.595 & --0.090    & 13  & 11  & 0.54 & 105\\
MWP1G012630--000100S$^{2}$ & 12.633 & --0.017    & 10  & 6   & 0.79 & 134\\
N8$^{3}$ & 12.805 & --0.312                      & 13  & 12  & 0.21 & 17\\
MWP1G013213--001410$^{2}$ & 13.213 & --0.141     & 30  & 21  & 0.73 & 60\\
N13$^{3}$ & 13.899 & --0.014                     & 33  & 24  & 0.68 & 170\\
N14$^{3}$ & 14.000 & --0.136                     & 78  & 71  & 0.41 & 80\\
G014.175+0.024$^{6,1}$ & 14.175 & 0.022          & 11  & 10  & 0.44 & 179\\
MWP1G014210--001100S$^{2}$ & 14.206 & --0.110    & 13  & 7   & 0.82 & 48\\
MWP1G014390--000200S$^{2}$ & 14.388 & --0.024    & 26  & 13  & 0.86 & 157\\
MWP1G014480--000000S$^{2}$ & 14.490 & 0.022      & 14  & 11  & 0.68 & 54\\
MWP1G016390--001400S$^{2}$ & 16.391 & --0.138    & 7   & 6   & 0.45 & 132\\
MWP1G016429--001984$^{2}$ & 16.431 & --0.201     & 57  & 44  & 0.64 & 14\\
MWP1G016560+000056$^{2}$ & 16.560 & 0.002        & 19  & 14  & 0.72 & 179\\
MWP1G017626+000493$^{2}$ & 17.625 & 0.048        & 24  & 23  & 0.29 & 70\\
TWKK1$^{4}$ & 17.805 & 0.074                     & 13  & 12  & 0.45 & 102\\
N20$^{3}$ & 17.918 & --0.687                     & 70  & 59  & 0.54 & 47\\
MWP1G018440+000100S$^{2}$ & 18.442 & 0.013       & 16  & 12  & 0.70  & 0\\
MWP1G018580+003400S$^{2}$ & 18.582 & 0.345       & 25  & 21  & 0.54  & 52\\
N23$^{3}$ & 18.679 & --0.237                     & 27  & 23  & 0.48 & 167\\
MWP1G018743+002521$^{2}$ & 18.748 & 0.256        & 31  & 25  & 0.58 & 51\\
MWP1G020387--000156$^{2}$ & 20.388 & --0.017     & 33  & 24  & 0.69 & 127\\
MWP1G02100--000500S$^{2}$ & 21.005 & --0.054     & 17  & 15  & 0.46 & 104\\
N28$^{3}$ & 21.351 & --0.137                     & 30  & 27  & 0.45 & 139\\
N31$^{3}$ & 23.842 & 0.098                       & 40  & 31  & 0.62 & 17\\
MWP1G023849--001251$^{2}$ & 23.848 & --0.127     & 20  & 17  & 0.46 & 146\\
MWP1G023881--003497$^{2}$ & 23.881 & --0.350     & 16  & 13  & 0.59 & 83\\
N32$^{3}$ & 23.904 & 0.070                       & 26  & 23  & 0.47 & 28\\
MWP1G023982--001096$^{2}$ & 23.982 & --0.110     & 19  & 16  & 0.54 & 85\\
MWP1G024019+001902$^{2}$ & 24.043 & 0.204        & 7   & 5   & 0.65 & 71\\
MWP1G024149--000060$^{2}$ & 24.153 & --0.011     & 12  & 6   & 0.88 & 16\\
N33$^{3}$ & 24.215 & --0.044                     & 26  & 22  & 0.55 & 24\\
TWKK3$^{4}$ & 24.424 & 0.220                     & 37  & 24  & 0.77 & 141\\
TWKK2$^{4}$ & 24.460 & 0.506                     & 12  & 6   & 0.87 & 49\\
MWP1G024500--002400$^{2}$ & 24.502 & --0.237     & 15  & 11  & 0.70 & 105\\
MWP1G024558--001329$^{2}$ & 24.558 & --0.133     & 50  & 43  & 0.53 & 104\\
MWP1G024649--001131$^{2}$ & 24.651 & --0.078     & 9   & 6   & 0.66 & 101\\
MWP1G01024699--001486$^{2}$ & 24.700 & --0.148   & 43  & 25  & 0.81 & 28\\
MWP1G024731+001580$^{2}$ & 24.736 & 0.158        & 29  & 25  & 0.52 & 49\\
MWP1G024920+000800$^{2}$ & 24.922 & 0.078        & 14  & 11  & 0.64 & 137\\
MWP1G025155+000609$^{2}$ & 25.155 & 0.061        & 25  & 19  & 0.66 & 141\\
MWP1G025723+00058$^{2}$ & 25.724 & 0.058         & 22  & 15  & 0.74 & 103\\
MWP1G025730--000200S$^{2}$ & 25.726 & --0.027    & 10  & 8   & 0.58 & 64\\
N42$^{3}$ & 26.329 & --0.071                     & 42  & 27  & 0.77 & 154\\
N43$^{3}$ & 26.595 & 0.095                       & 35  & 27  & 0.65 & 30\\
MWP1G026720+001700S$^{2}$ & 26.722 & 0.173       & 14  & 10  & 0.69 & 14\\
G027.492+0.192$^{6}$ & 27.496 & 0.197            & 29  & 27  & 0.38 & 108\\
MWP1G02671+00300S$^{2}$ & 27.613 & 0.028         & 14  & 11  & 0.59 & 117\\
MWP1G027905--000079$^{2}$ & 27.904 & --0.009     & 9   & 4   & 0.86 & 119\\
G027.9334+00.2056$^{6,5}$ & 27.931 & 0.205       & 14  & 9   & 0.74 & 142\\
MWP1G027981+000753$^{2}$ & 27.981 & 0.073        & 38  & 24  & 0.77 & 141\\
MWP1G028160--000300S$^{2}$ & 28.160 & --0.046    & 11  & 9   & 0.60 & 88\\
N49$^{3}$ & 28.827 & --0.229                     & 87  & 62  & 0.70 & 30\\
MWP1G029136--001438$^{2}$ & 29.134 & --0.144     & 27  & 16  & 0.80 & 86\\
N51$^{3}$ & 29.156 & --0.259                     & 130 & 109 & 0.54 & 46\\
MWP1G030020--000400S$^{2}$ & 30.022 & --0.041    & 19  & 17  & 0.47 & 16\\
MWP1G030250+002413$^{2}$ & 30.251 & 0.240        & 38  & 31  & 0.55 & 147\\
MWP1G03080+001100S$^{2}$ & 30.378 & 0.111        & 20  & 16  & 0.58 & 151\\
MWP1G030381--001074$^{2}$ & 30.381 & --0.109     & 17  & 12  & 0.69 & 141\\
MWP1G031066+000485$^{2}$ & 31.071 & 0.049        & 9   & 7   & 0.69 & 73\\
MWP1G032057+000783$^{2}$ & 32.055 & 0.076        & 49  & 42  & 0.52 & 44\\
N55$^{3}$ & 32.101 & 0.091                       & 65  & 43  & 0.75 & 170\\
MWP1G032731+002120$^{2}$ & 32.730 & 0.212        & 33  & 30  & 0.40 & 19\\
N57$^{3}$ & 32.761 & --0.149                     & 23  & 16  & 0.71 & 162\\
N60$^{3}$ & 33.815 & --0.149                     & 38  & 32  & 0.53 & 166\\
MWP1G034088+004405$^{2}$ & 34.087 & 0.441        & 41  & 22  & 0.84 & 142\\
MWP1G034680+000600S$^{2}$ & 34.684 & 0.067       & 12  & 8   & 0.70 & 102\\
N67$^{3}$ & 35.544 & 0.012                       & 52  & 43  & 0.57 & 10\\
MWP1G037196--004296$^{2}$ & 37.195 & --0.429     & 25  & 23  & 0.39 & 173\\
MWP1G037261--000809$^{2}$ & 37.258 & --0.078     & 26  & 20  & 0.65 & 74\\
MWP1G037349+006876$^{2}$ & 37.351 & 0.688        & 42  & 34  & 0.58 & 147\\
N70$^{3}$ & 37.750 & --0.113                     & 35  & 29  & 0.57 & 8\\
G038.550+1648$^{6}$ & 38.551 & 0.162             & 14  & 8   & 0.81 & 36\\
N73$^{3}$ & 38.736 & --0.140                     & 69  & 49  & 0.70 & 163\\
N78$^{3}$ & 41.228 & 0.169                       & 25  & 17  & 0.74 & 145\\
G041.378+0.035$^{6,1}$ & 41.378 & 0.034          & 10  & 8   & 0.54 & 43\\
N79$^{3}$ & 41.513 & 0.031                       & 81  & 53  & 0.76 & 103\\
TWKK4$^{4}$ & 41.595 & 0.160                     & 6   & 5   & 0.54 & 76\\
N80$^{3}$ & 41.932 & 0.033                       & 110 & 90  & 0.57 & 160\\
N89$^{3}$ & 43.739 & 0.114                       & 75  & 51  & 0.74 & 83\\
N90$^{3}$ & 43.774 & 0.060                       & 101 & 97  & 0.29 & 87\\
MWP1G045540+000000S$^{2}$ & 45.544 & --0.005     & 17  & 11  & 0.75 & 162\\
N96$^{3}$ & 46.949 & 0.371                       & 25  & 23  & 0.33 & 8\\
N98$^{3}$ & 47.027 & 0.218                       & 98  & 81  & 0.56 & 152\\
MWP1G048422+001173$^{2}$  & 48.422 & 0.116       & 35  & 30  & 0.55 & 123\\
N102$^{3}$ & 49.697 & --0.164                    & 122 & 112 & 0.39 & 138\\
N121$^{3}$ & 55.444 & 0.887                      & 36  & 33  & 0.43 & 46\\
\end{longtable}
}

\newpage
{\scriptsize
\begin{longtable}{|p{5cm}|p{2cm}|p{2cm}|p{1cm}|p{1cm}|p{1cm}|p{1cm}|}
\caption{Results of elliptical fits for the shapes of objects in 70 $\mu$m images. Objects were taken from: 1)Becker et al.~\cite{1994ApJS...91..347B}, 2) Simpson et al.~\cite{2012MNRAS.424.2442S}, 3) Churchwell et al.~\cite{2006ApJ...649..759C}, 4) new objects, 5) Urquhart et al.~\cite{2009A&A...507..795U}, 6) Egan et al.~\cite{2003yCat.5114....0E}.}\label{tab:catal70mkm}\\ \hline
Object & $l_{\rm gal}$, $^{\circ}$ & $b_{\rm gal}$, $^{\circ}$ & $a$, $^{''}$ & $b$, $^{''}$ & $e$ & PA, $^{\circ}$\\ \hline
\endfirsthead \hline
\multicolumn{7}{|c|}{\scriptsize\slshape(to be continued)} \\ \hline
Object & $l_{\rm gal}$, $^{\circ}$ & $b_{\rm gal}$, $^{\circ}$ & $a$, $^{''}$ & $b$, $^{''}$ & $e$ & PA, $^{\circ}$\\ \hline
\endhead \hline
\multicolumn{7}{|c|}{\scriptsize\slshape (to be continued)} \\ \hline
\endfoot \hline
\endlastfoot
S15$^{3}$ & 343.917 & --0.651 & 118 & 94 & 0.60 & 177\\
S21$^{3}$ & 341.357 & --0.288 & 50 & 38 & 0.65 & 179\\
S44$^{3}$ & 334.526 & 0.815 & 125 & 120 & 0.28 & 128\\
S123$^{3}$ & 312.979 & --0.434 & 128 & 123 & 0.27 & 160\\
S145$^{3}$ & 308.715 & 0.624 & 376 & 280 & 0.67 & 139\\
S167$^{3}$ & 301.630 & --0.349 & 394 & 305 & 0.63 & 117\\
CN67$^{3}$ & 5.525 & 0.037 & 42 & 33 & 0.62 & 58\\
CN77$^{3}$ & 6.140 & --0.640 & 68 & 49 & 0.69 & 74\\
CN79$^{3}$ & 6.203 & --0.335 & 71 & 41 & 0.82 & 115\\
CN111$^{3}$ & 8.306 & --0.084 & 89 & 74 & 0.55 & 122\\
MWP1G008430--002800S$^{2}$ & 8.431 & --0.274 & 22 & 12 & 0.82 & 45\\
CN116$^{3}$ & 8.476 & --0.278 & 16 & 11 & 0.72 & 107\\
N4$^{3}$ & 11.892 & 0.748 & 121 & 110 & 0.41 & 134\\
MWP1G012590--000900S$^{2}$ & --- & --- & --- & --- & --- & ---\\
MWP1G012630--000100S$^{2}$ & 12.633 & --0.017 & 7 & 7 & 0.24 & 2\\
N8 & 12.805 & --0.312 & 13 & 9 & 0.72 & 112\\
MWP1G013213--001410$^{2}$ & 13.212 & --0.138 & 26 & 19 & 0.68 & 22\\
N13$^{3}$ & 13.897 & --0.013 & 25 & 20 & 0.60 & 145\\
N14$^{3}$ & 14.002 & --0.135 & 83 & 61 & 0.68 & 91\\
G014.175+0.024$^{6,1}$ & 14.175 & 0.023 & 9 & 7 & 0.59 & 66\\
MWP1G014210--001100S$^{2}$ & --- & --- & --- & --- & --- & ---\\
MWP1G014390--000200S$^{2}$ & 14.390 & --0.023 & 13 & 12 & 0.49 & 16\\
MWP1G014480--000000S$^{2}$ & --- & --- & --- & --- & --- & ---\\
MWP1G016390--001400S$^{2}$ & --- & --- & --- & --- & --- & ---\\
MWP1G016429--001984$^{2}$ & 16.432 & --0.201 & 46 & 37 & 0.60 & 20\\
MWP1G016560+000056$^{2}$ & 16.559 & 0.002 & 14 & 13 & 0.42 & 2\\
MWP1G017626+000493$^{2}$ & 17.626 & 0.048 & 22 & 21 & 0.28 & 83\\
TWKK1$^{4}$ & 17.805 & 0.074 & 11 & 8 & 0.69 & 86\\
N20$^{3}$ & 17.917 & --0.684 & 58 & 53 & 0.41 & 180\\
MWP1G018440+000100S$^{2}$ & 18.442 & 0.013 & 11 & 6 & 0.84 & 151\\
MWP1G018580+003400S$^{2}$ & 18.583 & 0.347 & 21 & 16 & 0.68 & 87\\
N23$^{3}$ & 18.680 & --0.237 & 25 & 21 & 0.56 & 155\\
MWP1G018743+002521$^{2}$ & 18.748 & 0.257 & 27 & 25 & 0.39 & 27\\
MWP1G020387--000156$^{2}$ & 20.388 & --0.017 & 27 & 23 & 0.53 & 156\\
MWP1G02100--000500S$^{2}$ & 21.005 & --0.053 & 17 & 14 & 0.59 & 126\\
N28$^{3}$ & 21.351 & --0.138 & 31 & 26 & 0.55 & 149\\
N31$^{3}$ & 23.843 & 0.099 & 39 & 27 & 0.73 & 179\\
MWP1G023849--001251$^{2}$ & 23.849 & --0.128 & 18 & 16 & 0.40 & 17\\
MWP1G023881--003497$^{2}$ & 23.881 & --0.350 & 13 & 12 & 0.39 & 159\\
N32$^{3}$ & 23.903 & 0.069 & 25 & 20 & 0.61 & 113\\
MWP1G023982--001096$^{2}$ & 23.983 & --0.111 & 15 & 12 & 0.62 & 115\\
MWP1G024019+001902$^{2}$ & 24.043 & 0.204 & 7 & 7 & 0.43 & 134\\
MWP1G024149--000060$^{2}$ & 24.151 & --0.010 & 14 & 5 & 0.94 & 9\\
N33$^{3}$ & 24.215 & --0.044 & 24 & 18 & 0.68 & 179\\
TWKK3$^{4}$ & 24.427 & 0.222 & 25 & 21 & 0.52 & 75\\
TWKK2$^{4}$ & 24.461 & 0.507 & 11 & 7 & 0.79 & 37\\
MWP1G024500--002400$^{2}$ & 24.502 & --0.236 & 16 & 11 & 0.73 & 95\\
MWP1G024558--001329$^{2}$ & 24.558 & --0.134 & 44 & 37 & 0.55 & 61\\
MWP1G024649--001131$^{2}$ & --- & --- & --- & --- & --- & ---\\
MWP1G01024699--001486$^{2}$ & 24.699 & --0.147 & 43 & 27 & 0.78 & 12\\
MWP1G024731+001580$^{2}$ & 24.736 & 0.158 & 26 & 18 & 0.71 & 118\\
MWP1G024920+000800$^{2}$ & 24.922 & 0.079 & 10 & 9 & 0.43 & 7\\
MWP1G025155+000609$^{2}$ & 25.156 & 0.061 & 23 & 16 & 0.70 & 135\\
MWP1G025723+00058$^{2}$ & 25.724 & 0.058 & 19 & 13 & 0.74 & 118\\
MWP1G025730--000200S$^{2}$ & 25.726 & --0.027 & 9 & 7 & 0.63 & 62\\
N42$^{3}$ & 26.329 & --0.071 & 40 & 26 & 0.75 & 155\\
N43$^{3}$ & 26.594 & 0.096 & 32 & 26 & 0.57 & 46\\
MWP1G026720+001700S$^{2}$ & --- & --- & --- & --- & --- & ---\\
G027.492+0.192$^{6}$ & 27.496 & 0.197 & 30 & 21 & 0.70 & 101\\
MWP1G02671+00300S$^{2}$ & 27.613 & 0.028 & 11 & 8 & 0.71 & 158\\
MWP1G027905--000079$^{2}$ & --- & --- & --- & --- & --- & ---\\
G027.9334+00.2056$^{6,5}$ & 27.932 & 0.205 & 12 & 9 & 0.64 & 5\\
MWP1G027981+000753$^{2}$ & 27.979 & 0.072 & 26 & 23 & 0.48 & 71\\
MWP1G028160--000300S$^{2}$ & 28.160 & --0.046 & 11 & 10 & 0.42 & 147\\
N49$^{3}$ & 28.827 & --0.230 & 84 & 67 & 0.58 & 36\\
MWP1G029136--001438$^{2}$ & 29.133 & --0.146 & 19 & 15 & 0.58 & 46\\
N51$^{3}$ & 29.154 & --0.257 & 129 & 100 & 0.64 & 18\\
MWP1G030020--000400S$^{2}$ & 30.021 & --0.042 & 21 & 16 & 0.64 & 16\\
MWP1G030250+002413$^{2}$ & 30.251 & 0.240 & 25 & 19 & 0.67 & 8\\
MWP1G03080+001100S$^{2}$ & 30.378 & 0.110 & 19 & 16 & 0.54 & 134\\
MWP1G030381--0010744$^{2}$ & 30.381 & --0.109 & 15 & 13 & 0.40 & 151\\
MWP1G031066+000485$^{2}$ & --- & --- & --- & --- & --- & ---\\
MWP1G032057+000783$^{2}$ & 32.056 & 0.076 & 48 & 40 & 0.54 & 114\\
N55$^{3}$ & 32.100 & 0.091 & 59 & 42 & 0.71 & 163\\
MWP1G032731+002120$^{2}$ & 32.729 & 0.212 & 31 & 29 & 0.31 & 45\\
N57$^{3}$ & 32.762 & --0.149 & 20 & 16 & 0.60 & 166\\
N60$^{3}$ & 33.815 & --0.149 & 36 & 34 & 0.34 & 25\\
MWP1G034088+004405$^{2}$ & 34.086 & 0.441 & 28 & 22 & 0.65 & 145\\
MWP1G034680+000600S$^{2}$ & --- & --- & --- & --- & --- & ---\\
N67$^{3}$ & 35.544 & 0.012 & 49 & 40 & 0.57 & 22\\
MWP1G037196--004296$^{2}$ & 37.197 & --0.428 & 26 & 18 & 0.73 & 98\\
MWP1G037261--000809$^{2}$ & 37.258 & --0.078 & 22 & 18 & 0.57 & 80\\
MWP1G037349+006876$^{2}$ & 37.351 & 0.689 & 38 & 25 & 0.76 & 145\\
N70$^{3}$ & 37.749 & --0.113 & 34 & 25 & 0.66 & 158\\
G038.550+164$^{6}$ & 38.551 & 0.162 & 14 & 13 & 0.38 & 13\\
N73$^{3}$ & 38.738 & --0.139 & 54 & 51 & 0.32 & 55\\
N78$^{3}$ & 41.229 & 0.170 & 19 & 14 & 0.67 & 19\\
G041.378+0.035$^{6,1}$ & --- & --- & --- & --- & --- & ---\\
N79$^{3}$ & 41.514 & 0.030 & 74 & 59 & 0.60 & 107\\
TWKK4$^{4}$ & --- & --- & --- & --- & --- & ---\\
N80$^{3}$ & 41.937 & 0.032 & 99 & 71 & 0.70 & 136\\
N89$^{3}$ & 43.739 & 0.113 & 60 & 56 & 0.38 & 72\\
N90$^{3}$ & 43.774 & 0.060 & 101 & 96 & 0.31 & 94\\
MWP1G045540+000000SS$^{2}$ & --- & --- & --- & --- & --- & ---\\
N96$^{3}$ & 46.948 & 0.371 & 24 & 19 & 0.64 & 152\\
N98$^{3}$ & 47.029 & 0.219 & 87 & 82 & 0.34 & 148\\
MWP1G048422+001173$^{2}$  & 48.422 & 0.116 & 29 & 29 & 0.22 & 164\\
N102$^{3}$ & 49.698 & --0.163 & 117 & 112 & 0.29 & 136\\
N121$^{3}$ & 55.445 & 0.885 & 32 & 28 & 0.50 & 38\\
\end{longtable}                                                        
}                                                                      

\end{document}